\documentclass{revtex4-2}
\usepackage[utf8]{inputenc}
\usepackage{amsbsy}
\usepackage{amsopn}
\usepackage{amstext}
\usepackage{amsmath,amsthm,amsfonts,amssymb}
\usepackage[mathcal]{eucal}
\usepackage{mathrsfs}
\usepackage{braket}
\usepackage[english]{babel}
\usepackage{color}
\usepackage{esint}
\usepackage{graphicx}
\usepackage{float}
\usepackage{units}
\usepackage{blindtext}
\usepackage{hyperref}
\usepackage{textcomp}
\usepackage{orcidlink}
\usepackage{lineno}

\DeclareGraphicsExtensions{.png,PNG,.pdf,.PDF}
\usepackage{hyperref}
\usepackage{slashed}
\newcommand{\ie}{\begin{equation}}
\newcommand{\fe}{\end{equation}}
\newcommand{\se}{\begin{eqnarray}}
\newcommand{\ff}{\end{eqnarray}}
\begin{document}

\title{Comment on ``Quantum phase transitions of Dirac particles in a magnetized rotating curved background: Interplay of geometry, magnetization, and thermodynamics''}
\author{R. R. S. Oliveira\,\orcidlink{0000-0002-6346-0720}}
\email{rubensrso@fisica.ufc.br}
\affiliation{Departamento de F\'isica, Universidade Federal do Cear\'a (UFC), Campus do Pici, C.P. 6030, Fortaleza, CE, 60455-760, Brazil}


\date{\today}

\begin{abstract}
In this comment, we obtain the complete energy spectra for the paper by Sahan et al. \cite{Sahan}, that is, the energy spectra dependent on two quantum numbers, namely, the radial quantum number (given by $n\geq 0$) and the angular quantum number (given by $m\neq 0$). In particular, what motivated us to carry out such a study was the fact that the quantized energy spectra for Dirac particles in a curved or flat spacetime in polar coordinates explicitly depend on two quantum numbers. From this, the following question arose: Why do the energy spectra in the paper by Sahan et al. \cite{Sahan} depends on only one quantum number and not two, given that they worked with the Dirac equation in polar coordinates? So, using several important papers in the literature on the Dirac equation in curved spacetimes, as well as the most commonly used definition for Dirac gamma matrices in (2+1)-dimensions, we corrected some minor errors in the paper by Sahan et al. \cite{Sahan}. Consequently, we obtain the true second-order differential equation for their problem, as well as the complete energy spectra, which explicitly depend on both $n$ and $m$. Finally, we note that for $m<0$ (negative angular momentum) with $s=-1$ (lower component of the Dirac spinor), we obtain (except for one term with the incorrect sign) the particular energy spectra of Sahan et al. \cite{Sahan}.
\end{abstract}

\maketitle

\section{Introduction}

In a recent paper published in the Physics of the Dark Universe, entitled ``Quantum phase transitions of Dirac particles in a magnetized rotating curved background: Interplay of geometry, magnetization, and thermodynamics'', Sahan et al. \cite{Sahan} investigated the quantum and classical phase transitions of the Dirac particles in a homogeneously magnetized curved rotating 2+1-dimensional spacetime (i.e., Dirac particles in a uniform magnetic field in the Gürses spacetime). They considered the intricate relationship between geometry and quantum phase events through the study of quantum electrodynamics in the rotating curved spacetime. Using methods from quantum electrodynamics and statistical mechanics, they examined the effects of an external magnetic field, the background rotation parameter, and curvature on the characteristics of quantum and classical phase transitions, focusing on critical points and scaling behavior. So, in order to carry out all this study, they first had to solve the Dirac equation in the context of 2+1 gravity, that is, the Dirac equation written in curved spacetime for Dirac particles in a homogeneously magnetized 2+1-dimensional rotated curved spacetime (in polar coordinates). Having done that, they obtained what we can consider the basis (or the main result) of their paper, namely, the quantized energy spectra (for the system). In fact, these energy spectra were essential for Sahan et al. \cite{Sahan} to construct their partition function and, subsequently, obtain the thermodynamic properties, such as the magnetization, heat capacity, Helmholtz free energy, and the entropy, respectively. In particular, the paper by Sahan et al. \cite{Sahan} is well-written, is very interesting, and addresses an important subject involving Dirac particles in curved spacetimes. Furthermore, the formalism used by them (i.e., the spin connection or tetrad formalism) is also of great relevance in the literature when working with Dirac particles in curved spacetimes or in rotating frames of reference \cite{Alsing,Panahi,Chen,Ebihara,Chernodub,Bakke,Bakke2,Bragança,Andrade,Sadooghi,Kiessling,Hammad,Bueno,Neto,O1,O2,O3,O4,O5,O6,O7,O8,O9}.

However, starting from the fact that the quantized energy spectra for Dirac particles in a curved or flat spacetime in polar coordinates explicitly/intrinsically depend on two quantum numbers (a radial quantum number and an angular (momentum) quantum number) \cite{Bueno,Neto,O1,O2,O3,O4,O5,O6,O7,O8,O9,O10,Fabiano,Villalba1,Villalba2,Ferkous,Ghosh}, the following questions/doubts arose: Why do the energy spectra in the paper by Sahan et al. \cite{Sahan} depends on only one quantum number and not two, given that they worked with the Dirac equation in polar coordinates? Could these spectra be incomplete, since they should also depend on a second quantum number? In other words, where did the angular quantum number given by $m$ go? So, since Sahan et al. \cite{Sahan} did not explain why their spectra did not depend on the angular quantum number (since $m\neq 0$), this motivated us to try to understand why such a quantum number did not appear in the spectra, or better, to try to obtain the energy spectra in their complete/general form (i.e., dependent on the two quantum numbers, such as happens in Refs. \cite{Bueno,Neto,O1,O2,O3,O4,O5,O6,O7,O8,O9,O10,Fabiano,Villalba1,Villalba2,Ferkous,Ghosh}). Furthermore, it is very important to highlight that the thermodynamic properties derived from the incomplete spectra of Sahan et al. \cite{Sahan} is incorrect since the partition function must take into account the two quantum numbers in the spectra \cite{Pacheco,Arda,O11,O12,O13}. Therefore, the goal of the present comment is to obtain in detail the complete/general energy spectra for the paper by Sahan et al. \cite{Sahan}, that is, the energy spectra dependent on two quantum numbers: the radial quantum number and angular quantum number (as it should be). In particular, to achieve this goal, we will first correct some minor errors in both the curved Dirac equation and the second-order differential equation worked/used by Sahan et al. \cite{Sahan} in their paper. In addition, we will also obtain the differential equation for the upper component of the Dirac spinor, i.e., Sahan et al. \cite{Sahan} only obtained the equation for the lower component. Next, we will solve the corrected differential equation and obtain their respective energy spectra.


\section{Brief review of the main steps that Sahan et al. took to obtain their energy spectra}

According to Sahan et al. \cite{Sahan}, the Dirac equation in the context of 2+1 gravity is expressed as follows
\begin{equation}\label{1}
\left\{i\hbar\bar{\sigma}^{\mu}(x)
\left[\partial_{\mu}-\Gamma_{\mu}(x)
+ieA_{\mu}\right]\right\}\Psi(x)=m_e c^2\Psi(x),
\end{equation}
where $e$ and $m_e$ respectively signify the charge and mass of the electron, $A_\mu$ denotes the components of the electromagnetic potential, expressed as $A_\mu=(0,0,A_2)$ in which $A_2=-\frac{Br^2}{2}$ with $B$ representing the magnetic field, and the spin connection components are symbolized by $\Gamma_\mu (x)$.

According to Sahan et al. \cite{Sahan}, the generic metric expression of the perfect fluid sources in 2+1 dimensions spacetime can be defined as follows
\begin{equation}\label{2}
ds^2=(dt-\Omega(r)d\phi)^2dt^2-dr^2-r^2 d\phi^2,
\end{equation}
in which
\begin{equation}
\Omega(r)=-\frac{kr^2}{2},
\end{equation}
where $k$ is the vorticity parameter characterizing the spacetime, and keep in mind that the Gürses metrics. In particular, the choice $\Omega (r)$ corresponds to a Gödel-type rotation profile consistent with perfect fluid sources in 2+1 dimensions (we will use this information in the next section). So, considering the metric as emphasized in Eq. \eqref{2}, Sahan et al. \cite{Sahan} obtained the contravariant metric tensor $g^{\mu\nu}$ of this background in the following manner
\begin{equation}\label{3}
g^{\mu\nu}=\begin{pmatrix}
1-\frac{\Omega^2(r)}{r^2} & 0 & \frac{\Omega(r)}{r^2} \\
0 & -1 & 0 \\
\frac{\Omega(r)}{r^2} & 0 & -\frac{1}{r^2}
\end{pmatrix}.
\end{equation}

Additionally, the triads, denoted as $e^{a\nu}$ and related by $e^{a\mu}e_{a}{}^{\nu}=g^{\mu\nu}$,
are expressed by \cite{Sahan}
\begin{equation}\label{4}
e^{a}{}_{\nu} =
\begin{pmatrix}
1 & 0 & 0 \\
0 & -1 & 0 \\
\frac{\Omega(r)}{r} & 0 & -\frac{1}{r}
\end{pmatrix}.
\end{equation}

So, using \eqref{3} and \eqref{4}, Sahan et al. \cite{Sahan} obtained
\begin{align}\label{5}
& \Gamma_{0}=-\frac{\Omega'(r)}{8r}
\left[\sigma^{2},\sigma^{1}\right],
\nonumber\\
& \Gamma_{1}=\frac{i\,\Omega'(r)}{8r}
\left[\sigma^{2},\sigma^{3}\right],
\nonumber\\
& \Gamma_{2}=\frac{1}{4}\left\{\left(1-\frac{\Omega(r)\Omega'(r)}{2r}\right)\left[\sigma^{1},\sigma^{2}\right]+\frac{i\Omega'(r)}{2}\left[\sigma^{3},\sigma^{1}\right]\right\},
\end{align}
where $\sigma^1$, $\sigma^2$ and $\sigma^3$ are the Pauli matrices, and comes from $\tilde{\sigma}^a=(\sigma^3,i\sigma^1,i\sigma^2)$.

According to Sahan et al. \cite{Sahan}, in this case, the Dirac equation \eqref{1} becomes
\begin{equation}\label{6}
\left[-i\sigma^{3}\partial_{t}+\sigma^{1}\left(\partial_{r}+\frac{1}{2r}\right)+\sigma^{2}\left(\frac{\partial_{\phi}}{r}-\frac{\Omega(r)}{r}\partial_{t}-\frac{ieBr}{2}\right)-
\frac{\Omega'(r)}{4r}+m_e
\right]\Psi=0,
\end{equation}
or yet
\begin{equation}\label{7}
\left[-i\sigma^{3}E+\sigma^+\partial_-+\sigma^-\partial_+-
\frac{\Omega'(r)}{4r}+m_e
\right]\Psi=0,
\end{equation}
where $\sigma^\pm=\frac{\sigma^1\pm i \sigma^2}{2}$ and $\partial_\pm$ are given by
\begin{equation}\label{8}
\partial_\pm=\partial_{r}\pm i\frac{\partial_{\phi}}{r}\mp\frac{\Omega(r)}{r}E\pm m_e \omega_cr,
\end{equation}
being $\omega_c=\frac{eB}{2m_e}$ the cyclotron frequency.

So, using the form/representation of Pauli matrices, as well as the following two-component spinor
\begin{equation}\label{9}
\Psi=
\frac{e^{i(m\phi-Et)}}{\sqrt{r}}
\chi(r), \ \ \chi(r)=[\chi_+(r),\chi_-(r)]^T,
\end{equation}
Sahan et al. \cite{Sahan} transformed \eqref{7} in the following equation
\begin{equation}\label{10}
\begin{pmatrix}
-E-\dfrac{\Omega'(r)}{4r}+m_e & \partial_-\\
\partial_+ & E-\dfrac{\Omega'(r)}{4r}+m_e
\end{pmatrix}
\begin{pmatrix}
\chi_+(r) \\
\chi_-(r)
\end{pmatrix}= 0.
\end{equation}

So, using $\Omega=-\frac{kr^2}{2}$ in \eqref{10}, Sahan et al. \cite{Sahan} obtained two second-order differential equations (one for the spatial component of the spinor), given by
\begin{equation}\label{11}
\left[\frac{d^2}{dr^2}-\left(\frac{Ek}{2}+m_e \omega_c\right)^2r^2+\left(Ek+2m_e\omega_c\right)\left(m-\frac{1}{2}\right)-\lambda-\frac{m^2}{r^2}-\frac{m}{r^2}\right]\chi_{\pm}(r)=0,
\end{equation}
where $\lambda$ is an eigenvalue which satisfies $\partial_\mp\partial_\pm\chi_\pm=\lambda\chi_\pm$. Therefore, square-integrable radial spinors are \cite{Sahan}
\begin{equation}\label{12}
\chi_{\pm}(r)
=\frac{C_{\pm}
e^{-\frac{\left(E_k+2m_e\omega_c\right)r^2}{4}}}{\sqrt{r}}\left[\left(\frac{Ek+2m_e\omega_c}{2}\right)r^2\right]^{\frac{m}{2}+\frac{1}{4}}\times 
{}_1F_1\left(\frac{\lambda}{2Ek+4m_e \omega_c},m+\frac{1}{2},\frac{\left(Ek+2m_e \omega_c\right)r^2}{2}\right),
\end{equation}
where $C_\pm$ stand for arbitrary constants and, ${}_1F_1$ is the hypergeometric function. From the polynomial condition of ${}_1F_1$, it implies that $\frac{\lambda}{(2Ek+4m_e\omega_c)}=-n$ (quantization condition), with $n$ being a positive integer. Therefore, using this quantization condition, Sahan et al. \cite{Sahan} obtained/determined the energy spectra as follows
\begin{equation}\label{spectra1}
E_\pm=kn\pm\sqrt{
k^2 n^2+4m_e\omega_c n+m_e^2+\frac{m_e k}{2}+\frac{k^2}{16}}.
\end{equation}

From the above, we can clearly see that the energy spectra do not depend (explicitly or implicitly) on the angular quantum number $m$. In particular, this is quite strange, since, as the second-order differential equation \eqref{11} depends on it, one would expect that the spectra generated by this equation would also depend on it (unfortunately, this was not the case). In other words, Sahan et al. \cite{Sahan} solved the differential equation \eqref{11} incorrectly.


\section{The complete energy spectra for the paper by Sahan et al.}

According to Refs. \cite{Alsing,Panahi,Chen,Ebihara,Chernodub,Bakke,Bakke2,Bragança,Andrade,Sadooghi,Kiessling,Hammad,Bueno,Neto,O1,O2,O3,O4,O5,O6,O7,O8,O9,Lawrie}, the Dirac equation in the presence of an external electromagnetic field (or with minimal coupling) in a (3+1)-dimensional curved spacetime (or in the context of 3+1 gravity) is given by (in SI units)
\begin{equation}\label{14}
i\hbar\gamma^\mu (x)\left(\nabla_\mu (x)+\frac{iqA_\mu (x)}{\hbar}\right)\Psi=m_ec\Psi,
\end{equation}
where $\gamma^\mu (x)=e^{\mu}{}_{a}(x)\gamma^a$ are the curved/curvilinear gamma matrices, with $\gamma^a=(\gamma^0,\gamma^1,\gamma^2,\gamma^3)=(\gamma^0,\vec{\gamma})=(\beta,\beta\vec{\alpha})$ being the flat/Cartesian gamma matrices and $\beta$ and $\vec{\alpha}$ the original Dirac matrices (with $\vec{\alpha}=\beta\vec{\gamma}=$off-diag$(\vec{\sigma},\vec{\sigma})$ and $\vec{\sigma}=(\sigma^1,\sigma^2,\sigma^3)=(\sigma_x,\sigma_y,\sigma_z)$), $e^{\mu}{}_{a}(x)$ are the tetrads (tetrad field, vierbein vectors or vierbeins), $\nabla_\mu (x)=\partial_\mu+\Gamma_\mu (x)$ is the covariant derivative, being $\Gamma_\mu=-\frac{i}{4}\omega_{ab\mu}(x)\sigma^{ab}$ the spinorial connection (or spinor affine connection), $\omega_{ab\mu}(x)$ is the spin connection (an antisymmetric tensor in the indices $a$ and $b$, i.e., $\omega_{ab\mu}(x)=-\omega_{ba\mu}(x)$), $\sigma^{ab}=\frac{i}{2}[\gamma^a,\gamma^b]$ is an antisymmetric tensor (or ``second-rank Dirac tensor'', since this tensor is formed by $i$ times the product of the gamma matrices, i.e., $\sigma^{ab}=i\gamma^a\gamma^b=-i\gamma^b\gamma^a$ for $a\neq b$), $A_\mu (x)=e^{b}{}_{\mu}(x)A_b$ is the curved electromagnetic potential (curved four-potential or curved gauge field), being $A_b=(A_0,A_1,A_2,A_3)=(A_0,-\vec{A})$ the flat electromagnetic potential (flat four-potential or flat gauge field), $e^{b}{}_{\mu}(x)$ are the inverse tetrads, $\Psi$ the four-component Dirac spinor, and the (physical) quantities $q=\pm e$ and $m_e$ are the electric charge, and the rest mass of the Dirac fermion, being $\hbar$ and $c$ the reduced Planck constant and the speed of light. So, unlike \eqref{1}, \eqref{14} is dimensionally correct, meaning both sides have dimensions of linear momentum. That is, in \eqref{1}, the first side has the dimension of linear momentum (but the correct one would be $A_\mu/\hbar$) and the second has the dimension of energy (therefore, it is dimensionally incorrect).

Now, we need to write Eq. \eqref{14} in a (2+1)-dimensional curved spacetime. In particular, this is easily done by writing the matrices $\beta$ and $\vec{\alpha}$ directly in terms of the Pauli matrices, that is: $\beta=\sigma^3$ and $\vec{\alpha}=\vec{\sigma}=(\sigma^1,\sigma^2)$ \cite{Bueno,Andrade,O1,O2,O3,O4,O5,O6,Fabiano,Villalba1,Villalba2,Ferkous,Ghosh,O11,O12,O13,Bermudez1,Bermudez2,Mandal,Hagen1,Hagen2,Menculini,Gerbert}. Consequently, the gamma matrices $\gamma^a$ will be written as $\gamma^a=(\gamma^0,\gamma^1,\gamma^2)=(\sigma^3,\sigma^3\vec{\sigma})$, or better, $\gamma^a=(\sigma^3,\sigma^3\sigma^1,\sigma^3\sigma^2)=(\sigma^3,i\sigma^2,-i\sigma^1)$ \cite{Panahi,Bueno,Andrade,O1,O2,O3,O4,O5,O6,Fabiano,Villalba1,Villalba2,Ferkous,Ghosh,O11,O12,O13,Bermudez1,Bermudez2,Mandal,Hagen1,Hagen2,Menculini,Gerbert}, where we use the fact that the product of the Pauli matrices must satisfy the relation $\sigma^i\sigma^j=\delta^{ij}+i\epsilon^{ijk}\sigma^k$ ($i,j,k=1,2,3$) \cite{Alsing}. Furthermore (``a truth test these gamma matrices''), it is important to mention that in (1+1)-dimensions (for example, considering the $x$-axis), we have $\gamma^a=(\sigma^3,i\sigma^2)=(\sigma_z,i\sigma_y)$, that is, only the first two components of the gamma matrices are considered \cite{Shi,Pedernales}. In particular, the representation/form for the gamma matrices in (2+1)-dimensions is not unique, i.e., there are others, such as the one used by Sahan et al. \cite{Sahan}, which was also used masterfully in \cite{G1,G2,G3,G4}. In other words, regardless of the representation/form of the gamma matrices, the most important thing is that they obey/satisfy the Clifford algebra, i.e., the anticommutation relation of the gamma matrices (in fact, this is mandatory for the problem modeled by the Dirac equation to be mathematically consistent). It is also important to mention that working with the Dirac equation with only one spatial dimension does not necessarily mean that the gamma matrices must be written directly in terms of Pauli matrices. For example, when considering only the $z$-axis, $\gamma^0$ and $\gamma^3$ are still $4\times 4$ matrices (it seems that this is only valid for the $z$-axis) \cite{Blasone,Greiner,Leo1,Leo2}, that is, they are not written directly in terms of Pauli matrices as would happen if it were the $x$- or $y$-axis (in particular, the $z$-axis is an ``exception to the rule'', that is, $\gamma^0$ and $\gamma^3$ cannot be converted into a $2\times 2$ matrices).

Therefore, in (2+1)-dimensions, the correct form of the Dirac equation in a curved spacetime is given as follows
\begin{equation}\label{15}
i\hbar\bar{\sigma}^\mu (x)\left(\nabla_\mu (x)+\frac{iqA_\mu (x)}{\hbar}\right)\Psi=m_ec\Psi,
\end{equation}
where we define the curved sigma matrices (or the ``Dirac sigma matrices'' to differentiate from Pauli sigma matrices) as $\bar{\sigma}^\mu (x)=e^{\mu}{}_{a}(x)\tilde{\sigma}^a$, being $\tilde{\sigma}^a=(\sigma^3,i\sigma^2,-i\sigma^1)$, or in vector notation, such as $\tilde{\sigma}^a=(\sigma^3,i\vec{\sigma}_D)$, where $\vec{\sigma}_{D}=\vec{\sigma}_{Dirac}=(\sigma^2,-\sigma^1)$. Furthermore, the spinorial connection will now be written as $\Gamma_\mu=-\frac{i}{4}\omega_{ab\mu}\sigma^{ab}=-\frac{i}{4}\omega_{ab\mu}(\frac{i}{2}[\tilde{\sigma}^a,\tilde{\sigma}^b])=\frac{1}{8}\omega_{ab\mu}[\tilde{\sigma}^a,\tilde{\sigma}^b]$ (or $\Gamma_\mu=\frac{1}{4}\omega_{ab\mu}\tilde{\sigma}^a\tilde{\sigma}^b$ since $[\tilde{\sigma}^a,\tilde{\sigma}^b]=\tilde{\sigma}^a\tilde{\sigma}^b-\tilde{\sigma}^b\tilde{\sigma}^a=2\tilde{\sigma}^a\tilde{\sigma}^b$). Here, it is also important to highlight (which was not done by [1]) that since we are considering the polar coordinate system, we must have $\mu,\nu,\ldots=t,r,\phi$ (indices of curved spacetime) and $a,b,\ldots=0,1,2$ (indices of flat spacetime), respectively. 

So, using the same $\Omega(r)$ ($=-\frac{kr^2}{2}$) in Refs. \cite{O4,O5}, i.e., defining $\Omega\equiv\frac{k}{2}$ (was an unfortunate notation used by Ref. \cite{Sahan} since $\Omega$ was/is already used to symbolize/represent the vorticity (``rotation'') of the Gödel-type spacetime), the metric $g_{\mu\nu}(x)$ and its inverse $g^{\mu\nu}(x)$ are given as follows ($\hbar=c=1$)
\begin{equation}\label{16}
g_{\mu\nu}(x)=\begin{pmatrix}
1 & 0 & -\Omega(r) \\
0 & -1 & 0 \\
-\Omega(r) & 0 & \Omega^2(r)-r^{2}
\end{pmatrix}, \ \ g^{\mu\nu}(x)=\begin{pmatrix}
1-\frac{\Omega^2(r)}{r^2} & 0 & -\frac{\Omega(r)}{r^2} \\
0 & -1 & 0 \\
-\frac{\Omega(r)}{r^2} & 0 & -\dfrac{1}{r^{2}}
\end{pmatrix}.
\end{equation}

That is, besides Ref. \cite{Sahan} not showing $g_{\mu\nu}(x)$ in his paper, it made a sign error in two components of $g^{\mu\nu}(x)$, i.e., the component $g^{t\phi}=g^{\phi t}$ in this paper is given by $\frac{\Omega(r)}{r^2}$, while the correct value is given by $-\frac{\Omega(r)}{r^2}$ (as we can see above). We believe this error was caused by forgetting the negative sign in $\Omega (r)$. In fact, if this sign were not present, their inverse metric would be correct. In particular, the negative sign in $\Omega (r)$ is somewhat useless since the line element could simply have been written as $ds^2=(dt+\Omega(r)d\phi)^2-dr^2-r^2 d\phi^2$, where $\Omega (r)\equiv\frac{kr^2}{2}$ (unfortunately, that was not the case).

So, with the metrics \eqref{16} at hand, Refs. \cite{O4,O5} obtained the following tetrads and their inverses (unfortunately, Ref. \cite{Sahan} did not show this in his paper)
\begin{equation}\label{17}
e^{\mu}{}_{a}(x)=\begin{pmatrix}
1 & 0 & \frac{\Omega(r)}{r} \\
0 & 1 & 0 \\
0 & 0 & \frac{1}{r}
\end{pmatrix}, \ \ e^{a}{}_{\mu}(x)=\begin{pmatrix}
1 & 0 & -\Omega(r) \\
0 & 1 & 0 \\
0 & 0 & r
\end{pmatrix}.
\end{equation}

Consequently, the $2\times 2$ curved sigma matrices $\bar{\sigma}^\mu (x)=e^{\mu}{}_{a}(x)\tilde{\sigma}^a$ are written as \cite{O4,O5}
\begin{equation}\label{17}
\bar{\sigma}^{t}(x)=\tilde{\sigma}^{0}+\frac{\Omega (r)}{r}\tilde{\sigma}^{2},
\ \ \bar{\sigma}^{r}(x)=\tilde{\sigma}^{1},
\ \ \bar{\sigma}^{\phi}(x)=\frac{\tilde{\sigma}^{2}}{r}.
\end{equation}

Now, to have a Dirac equation written explicitly in terms of $\Omega(r)$, we need to know the form of the spinorial/spin connection contribution in the equation, that is, to know the form of $i\bar{\sigma}^\mu (x)\Gamma_\mu (x)$. So, known that $\Gamma_\mu (x)$ is given as follows \cite{O4,O5}
\begin{equation}\label{18}
\Gamma_{t}(x)=\frac{\Omega'(r)}{8r}[\tilde{\sigma}^{1},\tilde{\sigma}^{2}],
\ \ \Gamma_{r}(x)=\frac{\Omega'(r)}{8r}[\tilde{\sigma}^{0},\tilde{\sigma}^{2}], \ \ \Gamma_{\phi}(x)=\frac{1}{4}\left\{\left(1-\frac{\Omega(r)\Omega'(r)}{2r}\right)[\tilde{\sigma}^{1},\tilde{\sigma}^{2}]-\frac{\Omega'(r)}{2}[\tilde{\sigma}^{0},\tilde{\sigma}^{1}]\right\},
\end{equation}
we have
\begin{equation}\label{19}
[\bar{\sigma}^{t}(x)\Gamma_{t}(x)+\bar{\sigma}^{r}(x)\Gamma_{r}(x)+\bar{\sigma}^\phi(x)\Gamma_{\phi}(x)]=-\frac{\Omega'(r)}{4r}\tilde{\sigma}^{0}\tilde{\sigma}^{1}\tilde{\sigma}^{2}+\frac{\tilde{\sigma}^{1}}{2r},
\end{equation}
with
\begin{equation}\label{20}
i\bar{\sigma}^\mu(x)\Gamma_\mu (x)=-\frac{\Omega'(r)}{4r}i\tilde{\sigma}^{0}\tilde{\sigma}^{1}\tilde{\sigma}^{2}+\frac{i\tilde{\sigma}^{1}}{2r}=-\frac{\Omega'(r)}{4r}+\frac{i\tilde{\sigma}^{1}}{2r},
\end{equation}
where we use the fact that $i\tilde{\sigma}^{0}\tilde{\sigma}^{1}\tilde{\sigma}^{2}=i(\sigma^3)(i\sigma^2)(-i\sigma^1)=i(\sigma^3)(\sigma^2\sigma^1)=-i(\sigma^3)(\sigma^1\sigma^2)=-i(\sigma^3)(i\sigma^3)=(\sigma^3)^2=1$. In particular, comparing \eqref{18} with \eqref{5}, we see that Sahan et al. \cite{Sahan} made mistakes in some signs, as well as in some commutators, since $[\tilde{\sigma}^{1},\tilde{\sigma}^{2}]=[i\sigma^{2},-i\sigma^{1}]=[\sigma^2,\sigma^1]=-[\sigma^1,\sigma^2]$, $[\tilde{\sigma}^{0},\tilde{\sigma}^{2}]=[\sigma^3,-i\sigma^1]=-i[\sigma^3,\sigma^1]=i[\sigma^1,\sigma^3]$, and $[\tilde{\sigma}^{0},\tilde{\sigma}^{1}]=[\sigma^3,i\sigma^2]=i[\sigma^3,\sigma^2]$. That is, in \eqref{5}, $\Gamma_0$ is incorrect in the sign, $\Gamma_1$ is incorrect in the commutator, and $\Gamma_2$ is incorrect in the sign with respect to the first term and in both the sign and commutator in the second term, respectively. In fact, we believe this is due to the definition of $\tilde{\sigma}^a$, which Sahan et al. \cite{Sahan} did not follow the standard definition/prescription used in the literature (which, as we have seen, begins by writing $\beta$ and $\vec{\alpha}$ directly in terms of Pauli matrices).

Therefore, using \eqref{17}, and \eqref{20}, with $A_\mu (x)=(0,0,A_\phi (x))=(0,0,e^2_\phi (x)A_2)=(0,0,-rA_\phi(r))=(0,0,-\frac{Br^2}{2})$ \cite{O2,O3,O4,O6,O7} (i.e., this potential has the same form as \cite{Sahan}) and also $\hbar=c=1$ (natural units), into \eqref{15}, we obtain
\begin{equation}\label{21}
\left[i\tilde{\sigma}^{0}\partial_t+i\tilde{\sigma}^{1}\left(\partial_r+\frac{1}{2r}\right)+\tilde{\sigma}^{2}\left(\frac{1}{r}i\partial_\phi+\frac{\Omega (r)}{r}i\partial_t+\frac{qBr}{2}\right)-\frac{\Omega'(r)}{4r}-m_e\right]\Psi=0,
\end{equation}
or better (multiplying everything by $-1$ and writing in terms of $\sigma^1$, $\sigma^2$, and $\sigma^3$)
\begin{equation}\label{22}
\left[-i\sigma^3\partial_t+\sigma^{2}\left(\partial_r+\frac{1}{2r}\right)+i\sigma^1\left(\frac{1}{r}i\partial_\phi+\frac{\Omega (r)}{r}i\partial_t+\frac{qBr}{2}\right)+\frac{\Omega'(r)}{4r}+m_e\right]\Psi=0,
\end{equation}
or yet (with $q=-e$, i.e., considering the electron)
\begin{equation}\label{23}
\left[-i\sigma^3\partial_t+\sigma^{2}\left(\partial_r+\frac{1}{2r}\right)-\sigma^1\left(\frac{1}{r}\partial_\phi+\frac{\Omega (r)}{r}\partial_t+\frac{ieBr}{2}\right)+\frac{\Omega'(r)}{4r}+m_e\right]\Psi=0.
\end{equation}

In particular, comparing \eqref{23} with \eqref{6}, we see that only the first and last terms are the same. However, unlike \eqref{6}, in \eqref{23}, the second matrix is $\sigma^2$, the third matrix is $\sigma^1$, the term associated with the magnetic field is positive, and the penultimate term is positive.

So, using the spinor \eqref{9} (but without $\sqrt{r}$ and with $\chi(r)=[\chi_+(r),i\chi_-(r)]^T$, such as is done in Refs. \cite{O4,O5,Fabiano}), as well as $\Omega(r)=-\frac{kr^2}{2}$, we can obtain from \eqref{23} the following radial Dirac equation (here, the correct way is to use $\frac{d}{dr}$ and not $\partial_r$ anymore)
\begin{equation}\label{24}
\left[-\sigma^3 E+\sigma^{2}\left(\frac{d}{dr}+\frac{1}{2r}\right)-i\sigma^1\left(\frac{m}{r}+\frac{(Ek+eB)}{2}r\right)+\left(m_e-\frac{k}{4}\right)\right]\chi(r)=0,
\end{equation}
or better
\begin{equation}\label{25}
\left[-\sigma^3 E+\sigma^{2}\left(\frac{d}{dr}+\frac{1}{2r}\right)-i\sigma^1\left(\frac{m}{r}+\frac{(Ek+m\omega_c)}{2}r\right)+\left(m_e-\frac{k}{4}\right)\right]\chi(r)=0,
\end{equation}
where $\omega_c=\frac{eB}{m}>0$ is the famous cyclotron frequency \cite{O2,O3,O4,O6,O7} (the correct is without the factor 2, that is, not as it was used in \cite{Sahan}), and $m$ is actually the total magnetic quantum number $m_j\neq 0$, that is, $m=m_j=k=\pm\frac{1}{2},\pm\frac{3}{2},\pm\frac{5}{2}\ldots=l+\frac{1}{2}=m_l+m_s$ \cite{Panahi,Bakke,Bakke2,Bueno,O1,O2,O3,O4,O5,O6,O7,O8,O9,O10,O11,O12,O13,Villalba1,Villalba2}. Besides, using the form/representation of Pauli matrices, Eq. \eqref{25} becomes
\begin{equation}\label{26}
\begin{pmatrix}
-E+\left(m_e-\frac{k}{4}\right)\
& -i\left(\frac{d}{dr}+\frac{1}{2r}\right)
-i\left(\frac{m}{r}+\frac{(Ek+m_e\omega_c)}{2}r\right)
\\ i\left(\frac{d}{dr}+\frac{1}{2r}\right)
-i\left(\frac{m}{r}+\frac{(Ek+m_e\omega_c)}{2}r\right)
& E+\left(m_e-\frac{k}{4}\right)
\end{pmatrix}
\begin{pmatrix}
\chi_+(r)\\
\chi_-(r)
\end{pmatrix}=0.
\end{equation}
or better
\begin{equation}\label{27}
\begin{pmatrix}
-\left[E-\left(m_e-\frac{k}{4}\right)\right]
& -i\left[\frac{d}{dr}+\frac{(Ek+m_e\omega_c)}{2}r+\frac{(m+\frac{1}{2})}{r}\right]
\\
i\left[\frac{d}{dr}-\frac{(Ek+m_e\omega_c)}{2}r-\frac{(m-\frac{1}{2})}{r}\right]
& \left[E+\left(m_e-\frac{k}{4}\right)\right]
\end{pmatrix}
\begin{pmatrix}
\chi_+(r)\\
i\chi_-(r)
\end{pmatrix}=0.
\end{equation}

Therefore, from \eqref{27}, we obtain two coupled first-order differential equations, given by
\begin{align}\label{28}
& \left[\frac{d}{dr}+\frac{(Ek+m_e\omega_c)}{2}r+\frac{(m+\frac{1}{2})}{r}\right]\chi_-(r)-\left[E-\left(m_e-\frac{k}{4}\right)\right]\chi_+(r)=0,
\\\label{29}
& \left[\frac{d}{dr}-\frac{(Ek+m_e\omega_c)}{2}r-\frac{(m-\frac{1}{2})}{r}\right]\chi_+(r)+\left[E+\left(m_e-\frac{k}{4}\right)\right]\chi_-(r)=0.
\end{align}

Now, we need to obtain a second-order equation from the equations above; that is, we need to decouple these equations. In particular, this is done by isolating one of the components in one equation and substituting it into the other equation. Therefore, isolating $\chi_-(r)$ in \eqref{29} and substituting in \eqref{28}, and then isolating $\chi_+(r)$ in \eqref{28} and substituting in \eqref{29}, we obtain a second-order differential equation for each component. Then, in a compact form (i.e., writing ``two equations in just one''), we have the following second-order differential equation for the two components
\begin{equation}\label{30}
\left[\frac{d^2}{dr^2}+\frac{1}{r}\frac{d}{dr}-\frac{\left(m-\frac{s}{2}\right)^2}{r^2}-\frac{(Ek+m_e\omega_c)^2}{4}r^2+E^2-\left(m_e-\frac{k}{4}\right)^2-(Ek+m_e\omega_c)\left(m+\frac{s}{2}\right)\right]
\chi_s(r)=0,
\end{equation}
where $s$ is a real parameter whose values are $\pm 1$ (or $\pm$), being $s=+1$ for the for upper component ($\chi_+ (r)=\chi_{upper}(r)$), and $s=-1$ for the lower component ($\chi_- (r)=\chi_{lower}(r)$), respectively. So, unlike \cite{Sahan}, here, we have the differential equation for both components of the spinor. Furthermore, we could easily have used the factor $\sqrt{r}$ in the spinor (such as is done in \cite{O6} as well as in \cite{Villalba1,Villalba2}), however, to directly solve Eq. \eqref{30} and consequently to have both positive and negative values of $m$ (embedded) in each component, we prefer to omit this factor (i.e., it is optional). In other words, if we had used the factor, we would have two expressions for each component, where one would be for $ms>0$, and another for $ms<0$, respectively. Therefore, we omit the factor to make it easier for us to solve Eq. \eqref{30}.

So, to solve exactly/analytically Eq. \eqref{30}, let us first introduce a new (dimensionless real) variable, given by: $\rho=\frac{(Ek+m_e \omega_c)}{2}r^2$ (with $(Ek+m_e \omega_c)>0$ or $m_e \omega_c>-Ek$). In this way, through a change of variable, we have
\begin{equation}\label{31}
\left\{\rho\frac{d^2}{d\rho^2}
+\frac{d}{d\rho}-\frac{\left(m-\frac{s}{2}\right)^2}{4\rho}-\frac{\rho}{4}+\frac{1}{2(Ek+m_e\omega_c)}
\Big[E^2-\left(m_e-\frac{k}{4}\right)^2
-(Ek+m_e\omega_c)\left(m+\frac{s}{2}\right)\Big]\right\}
\chi_s(\rho)=0.
\end{equation}

Now, analyzing the asymptotic behavior/limit of Eq. \eqref{31} for $\rho\to 0$ and $\rho\to\infty$, we have a regular solution to this equation, given by
\begin{equation}\label{32}
\chi_s (\rho)=c_s\rho^{\frac{\vert m-\frac{s}{2}\vert}{2}}e^{-\frac{\rho}{2}}F_s (\rho),
\end{equation}
where $c_s>0$ are normalization constants, $F_s (\rho)$ are unknown functions to be determined.

So, substituting \eqref{32} in \eqref{31}, we have a second-order differential equation for $F_s (\rho)$, given by
\begin{equation}\label{33}
\rho F''_s(\rho)+(b-\rho)F'_s (\rho)-aF_s (\rho)=0,
\end{equation}
where we define
\begin{equation}\label{34}
b=b_s\equiv\Big\vert m-\frac{s}{2}\Big\vert+1, \ \ a=a_s\equiv\frac{\Big\vert m-\frac{s}{2}\Big\vert+1}{2}-\frac{1}{2(Ek+m_e\omega_c)}
\Big[E^2-\left(m_e-\frac{k}{4}\right)^2
-(Ek+m_e\omega_c)\left(m+\frac{s}{2}\right)\Big]. 
\end{equation}

According to the literature \cite{Bueno,Andrade,Fabiano,Bragança}, Eq. \eqref{33} is the well-known/famous confluent hypergeometric differential equation, or simply, the confluent hypergeometric equation (which can also be written in terms of the associated/generated Laguerre polynomials \cite{Andrade,Bragança,O4,O5,O6}), whose (only acceptable) solution this equation is the confluent hypergeometric function of the first kind, given by $F_s(\rho)={}_1F_1(a,b,\rho)$. So, for the Dirac spinor to be a normalizable (square-integrable) or physically consistent function (to describe the bound states of the fermion), it is necessary that $\chi_s (\rho)$ tend to zero when $\rho\to 0$ or $\rho\to\infty$ (which are the two boundary conditions of the system). Consequently, $F_s(\rho)$ must be a finite confluent hypergeometric series, which implies that the independent term in \eqref{33} is a negative integer (or zero), that is, $a=-n$ ($n=0,1,2,\ldots$) \cite{Bueno,Andrade,Fabiano,Bragança,O4,O5,O6}. In other words, we have a quantization condition for the system, where $n$ is often called the radial quantum number (since it arises from a second-order radial differential equation). Therefore, from this quantization condition (i.e., $a=-n$), we obtain the following energy spectra (or complete energy spectra for the paper by Sahan et al. \cite{Sahan})
\begin{equation}\label{spectra2}
E^{\pm}=E^\pm_{n,m,u}=kN\pm\sqrt{k^2 N^2+2m_e\omega_c N+\left(m_e-\frac{k}{4}\right)^2}=kN\pm\sqrt{k^2 N^2+2m_e\omega_c N+m_e^2-\frac{m_e k}{2}+\frac{k^2}{16}},
\end{equation}
where we define
\begin{equation}
N=N_{n,m,s}\equiv\left(n+\frac{1}{2}+\frac{\vert m-\frac{s}{2}\vert+\left(m+\frac{s}{2}\right)}{2}\right)=\left(n+\frac{1+s}{2}+\frac{\vert m-\frac{s}{2}\vert+\left(m-\frac{s}{2}\right)}{2}\right)\geq 0,
\end{equation}
being $N=N_{\mathbf{eff}}=N_{total}$ an ``effective quantum number'' or ``total quantum number'' (since it depends on all others). In particular, comparing \eqref{spectra2} with \eqref{spectra1}, we see that the difference between them is with respect to the quantum numbers, which here clearly/explicitly depend on the angular quantum number $m$ (and also of the parameter $s$ that describes the two components of the spinor), as well as with respect to the penultimate term of the square root, which here is negative (i.e., the negative sign, as well as the definition of $\omega_c$ are, basically, two ``small corrections'' for the energy spectra \eqref{spectra1}). So, for the energy spectra \eqref{spectra2} depend only on the radial quantum number $n$ (i.e., $N=n$), it is necessary that $m$ be negative ($m<0$), and consider the lower component ($s=-1$). In this way (and doing $\omega_c\to 2\omega_c$, with $\omega_c=\frac{eB}{2m}$, and $-\frac{m_e k}{2}\to +\frac{m_e k}{2}$), we then obtain the particular energy spectra of Sahan et al. \cite{Sahan}.


\section{Final remarks}

In this comment, we obtain the complete energy spectra for the paper by Sahan et al. \cite{Sahan}, that is, the energy spectra dependent on two quantum numbers, namely, the radial quantum number (given by $n\geq 0$) and the angular quantum number (given by $m\neq 0$). In particular, what motivated us to carry out such a study was the fact that the quantized energy spectra for Dirac particles (or Dirac fermions) in a curved or flat spacetime in polar coordinates explicitly depend on two quantum numbers (radial quantum number and the angular quantum number). From this, the following questions/doubts arose: Why do the energy spectra in the paper by Sahan et al. \cite{Sahan} depends on only one quantum number and not two, given that they worked with the Dirac equation in polar coordinates? Could these spectra be incomplete, since they should also depend on a second quantum number? In other words, where did the angular quantum number given by $m$ go, since $m\neq 0$? So, using several important papers in the literature on the Dirac equation in curved spacetimes, as well as the most commonly used definition for Dirac gamma matrices in (2+1)-dimensions, we corrected some minor errors in the paper by Sahan et al. \cite{Sahan}. Consequently, we obtain the true second-order differential equation for their problem, as well as the complete energy spectra, which explicitly/intrinsically depend on both $n$ and $m$. Finally, we note that for $m<0$ (negative angular momentum) with $s=-1$ (lower component of the Dirac spinor), we obtain (except for one term with the incorrect sign) the particular energy spectra of Sahan et al. \cite{Sahan}.
 
\section*{Acknowledgments}

\hspace{0.5cm}The author would like to thank the anonymous reviewer for their careful reading of the article, which contributed substantially to improving its quality, as well as the Conselho Nacional de Desenvolvimento Cient\'{\i}fico e Tecnol\'{o}gico (CNPq) for financial support.

\section*{Data availability statement}

\hspace{0.5cm} This manuscript has no associated data or the data will not be deposited.

\end{document}